\documentstyle[11pt]{article}
\date{}
\begin{document}
\sloppy
\title{Is the Energy Density of the Cosmic Quaternionic Field a Possible
Candidate for the Black Energy?}
\author{V. Majern\'{\i}k \\
Institute of Mathematics, Slovak
Academy of Sciences, \\ Bratislava, \v Stef\'anikova  47,
 Slovak Republic\\ and\\
Department of Theoretical Physics\\ Palack\'y
University\\T\v r. 17. listopadu 50\\772 07 Olomouc, Czech Republic}
\maketitle
\begin{abstract}
We try to show that the energy density of the cosmic quaternionic field
might be a possible candidate for the black energy.
\end{abstract}
\section{Introduction}
According to the astronomical observation, the evidence continues to mount
that the expansion of the
universe is accelerating  rather than slowing down.
New observation suggests a
universe that is leigh-weight, is accelerating, and is flat
\cite{PER} \cite{pe} \cite{BO}.
To induce cosmic acceleration
it is necessary to consider some
components, whose equations of state are different from baryons,
neutrinos, dark matter, or radiation considered in the standard
cosmology.

As is well-known, in cosmology a new kind of energy is considered
called {\it quintessence} ("dark energy"). Quintessence represents
a dynamical
form of energy with negative pressure \cite{ST}.
The quintessence is supposed to obey an equation of state of the form
\begin{equation} \label{1}
p_Qc^{-2}=w_Q\varrho_Q,\qquad
-1<w_Q<0.
\end{equation}
For the vacuum energy
(static cosmological
constant), it holds $w_Q=-1$ and $\dot w_Q=0$.

An adequate theory of a cosmological scenario, conform with
recent observation, should give answers to
the following problems:\\
(i){\it The cosmological constant problem.}
The '$\Lambda$-problem' can be expressed as discrepancies
between the negligible value of $\Lambda$ for the present universe and
the value $10^{50}$ times larger expected by Glashow-Salam-Weinberg model
\cite{WM}
or by GUT \cite{GUT} where it should be $10^{107}$ times larger.\\
{\it The fine-tuning problem}.
Assuming that the vacuum energy density is constant over time and the matter
density decreases as the universe expands it appears that their ratio
must be set to immense small value ($\approx 10^{-120}$) in the early
universe in order for the two densities to nearly {\it coincide} today, some
billions years later.\\
(iii) {\it The age problem}.
This problem expresses the discrepancy connected,
 on the one side,
with the hight estimates of the Hubble parameter and with the age of globular
clusters on the other side. The fact that the age of the universe is
smaller than the age of globular clusters is unacceptable.\\
(iv) {\it The flatness problem}.
Inflation predicts a spatially flat universe. According to Einstein's
theory, the mean energy density determines the spatial curvature of the
universe. For a flat universe, it must be equal to the critical energy.
The observed energy density is about one-third of critical
density. The discrepancy between the value of the observed energy
density and the critical energy is called the flatness problem.\\
(v) {\it Problem of the particle creation}.
In variable $\lambda$ models the creation of particles generally
takes place. The question what is the mechanism for this process
represents the problem of the particle creation.

It is well-known that the Einstein field equations with
a non-zero $\lambda$ can be rearranged so
that their right-hand sides consist of two terms:
the stress-energy tensor of the ordinary
matter and an additional tensor
$$T^{(\nu)}_{ij}=\left (\frac{c^4\lambda}{8\pi G}\right)
g_{ij}=\Lambda g_{ij}. \quad \eqno(2)$$
$\Lambda$ is identified with vacuum energy
because this quantity satisfies the requirements asked from $\Lambda$, i.e.
(i) it should have the dimension of energy density, and (ii) it
should be invariant under Lorentz transformation. The second property
is not satisfied for arbitrary systems, e.g. material systems and
radiation. Gliner \cite{G} has shown that the
energy density of vacuum represents a scalar function of the
four-dimensional space-time coordinates so that it satisfies both above
requirements. This is why  $\Lambda$ is {\it identified}
with the vacuum energy.\\
From what has been said so far it follows that
the following properties are required from the vacuum energy density:
(i) It should be intrinsically relativistic quantity having the dimension
of the energy density.
(ii) It should be smoothly distributed throughout the universe.
(iii) It should cause the speedup of the universe.
(iv) It should balances the total mean energy density to $\Omega=1$.

In the next Sections we will describe a model of the universe consisting
of a mixture of the ordinary matter and a so-called cosmic quaternionic
field \cite{MK} whose
energy density we set equal to the
cosmological constant. We show that the value of the energy density of
cosmic quaternionic field is consistent with the data. Then, we describe
the force exerting on the moving bodies in the cosmic quaternionic
field and the possible mechanism of the particle creation in this field.
Finally, we sketch the evolution of the universe with the cosmic
quaternionic field.

\section{The cosmic quaternionic field}

In a very recent article \cite{MK},
$\Lambda$ has been interpreted as the {\it field energy} of a  classical
quaternionic field (called $\Phi$-field, for short)
\cite{MM} \cite{SIG} \cite{A} which is
given by the field tensor $F_{ij}\quad i,j=1,2,3,0$ whose
components are defined as
$F_{ij}=0$ for $i \neq j$ and $F_{11}=F_{22}=F_{33}=-F_{00}=\Phi.$
The $\Phi$-field belongs to the family of the quaternionic
fields (see \cite{MM}).
The quaternionic field which we consider is given
by the field tensor which, in the matrix, has the form
$$F_{ij}=
\left(\begin{array}{cccc}
\Phi&0&0&0\\
0&\Phi&0&0\\
0&0&\Phi&0\\
0&0&0&-\Phi
\end{array} \right).$$
$\Phi$ is the only field variable in it.
$F_{ij}$ is a symmetric field tensor with
the components $F_{ii}=\Phi\quad i=1,2,3,\quad F_{ii}=-\Phi\quad
i=0,$ and $F_{ij}=0\quad
i\neq j.$  It is easily to show that $\Phi$ is transformed as a scalar
under Lorentz
transformation \cite{MK}.
The field equations of the $\Phi$-field in the differential are
$$\nabla \Phi=k \vec J\quad\quad i=1,2,3\quad\quad
{\rm and}\qquad
-{1\over c}{\partial \Phi \over \partial t}=k_{0}J_0,\quad
\quad
i=0,\quad \eqno(3)$$
where
$$k=\frac{\sqrt{G}}{4\pi c}
\quad{\rm and}\quad k_0=8\pi\sqrt{G}.$$
These equations are first-order differential equations whose
solution can be found given the source terms.
Assuming the spacial homogeneity of the $\Phi$-field it
becomes independent of spatial
coordinates therefore
it holds $J_1= J_2= J_3=0.$ The source of the $\Phi$-field is its {\it
own} mass density associated with the field energy density, i.e. $\Phi^2/8\pi
c^2$, therefore, it holds
$J_0=\Phi^2/8\pi
c^2$. $J_0$ is dependent only on time.
The energy
density associated with the field
is \cite{MM}
$${E_{\Phi}} ={\Phi^2 \over 8\pi}.$$
Since the current 4-vector in the everywhere local rest
frame has only one non-zero component, $J_0$,
 Eqs.(3) become
$$\nabla\Phi= 0$$
$$-{1\over c}{d \Phi \over d t}= {8\pi\sqrt {G}\Phi^2\over
8\pi c^2} ={\sqrt {G}\Phi^2 \over c^2}$$
whose solution is
$$\Phi(t)= {c \over \sqrt {G} (t+t_0)},\quad\eqno(4)$$
where $t_0$ is the integration constant given by the boundary condition.
The energy density of the $\Phi$-field $E_{\Phi}$ is approximately equal
to the observed value of the cosmological constant.

\section{The black energy modeled by a time-dependent cosmological constant}

Next, we set
$\Lambda$ equal to the {\it field energy density} of the cosmic
quaternionic field ($t_0=c=1$)
$$\Lambda=\frac{1}{8\pi }\left [\frac{1}{\sqrt{G}t}
\frac{1}{\sqrt{G}t}\right ]
=\frac{\Phi^{2}}{8\pi}\qquad \Phi=\frac{1}{\sqrt{G}t},
$$
Accordingly, we have
$$\Lambda =\frac{\Phi^2}{8\pi}=\frac{1}{8\pi G t^2}\qquad {\rm
and}\qquad \lambda=\frac{1}{t^2}.\quad \eqno(5)$$
The gravitational field equations with a
cosmological constant $\lambda$ and the
energy conservation law are (k=0)
$$H^2=\frac{8\pi G}{3}(\rho+\Lambda)\qquad H=\frac{\dot R}{R}\qquad
\Lambda=\frac{\lambda}{8\pi G}\quad\eqno(6)$$
$$\frac{\ddot R}{R}=-\frac{4\pi G}{3}(\rho+3p+2\Lambda)\quad\eqno(7)$$
and
$$\dot \rho+3\frac{\dot R}{R}(p+\rho)=-\dot \Lambda.\quad \eqno(8)$$
Suppose we have a perfect-gas equation of state
$$p=\alpha \rho\quad \eqno(9)$$
and suppose that the deceleration parameter is constant.
If the evolution of the scale factor is given in form $R\propto t^n$ then
$q=-(n-1)/n$, therefore, we set
$$q=-\frac{\ddot RR}{\dot R^2}=\frac{1}{n}-1.\quad\eqno(10)$$
If we suppose the time dependence $\rho$ and $\Lambda$ in form
$$\rho=\frac{A}{t^2}\qquad {\rm and} \qquad \Lambda=\frac{B}{t^2}\qquad
B=const. \qquad {\rm and}\qquad A=const., \quad
\eqno(11)$$
respectively, then, inserting (11) into (6),(7) and (8), gives the
following relation between $A$ and
$B$ \cite{AB}
$$2B=A[(-2+3n)(1+\alpha)].\quad \eqno(12)$$
Given $A$ or $B$ and $n$ we can uniquely determine $B$ or $A$,
respectively.
For $\lambda \propto 1/t^2$, there is a relation between $\Omega_M$ and
the time-dependence of
scaling factor $R(t)$. Assuming that $\Omega_M$ does not change
during the matter-dominated era ($\alpha=0$) \cite{KM}
$$R(t)=\left (\frac {3}{2}\right)^{\frac{2}{3\Omega_M}}(\Omega_M
C_1t)^{\frac{2}{3\Omega_M}}.\quad\eqno(13)$$
The quantities $q$, $R(t)$ and $\Omega_M$ are mutually related.
Given one of them the remained quantities can be determined by means of
Eqs. (13), (10) and (11). It seem that $\Omega_M$ is best determined
by the observation,
therefore, we take it for the calculation of $q$ and $R(t)$.
Inserting $\Omega_M=1/3$ into Eq.(13) we obtain
$R\propto t^2 $ which yields $q=-1/2=1/n-1$, $n=2$.
We see from Eq.(5) that $B=1$ which inserting into Eq.(12) gives
$A= (1/2)$. The mean energy density $\rho$ and the cosmological
constant is given as ($\alpha=0$)
$$\rho = \frac{1}{16\pi t^2}\qquad {\rm and}\qquad \lambda=\frac{1}{8\pi
G t^2}, \quad \eqno (14)$$
respectively.
Their ratio
$$\frac{\rho}{\Lambda}=1/2.$$
Supposing the flat space, we have
$$\Omega_M=\frac{1}{3}\qquad {\rm and}\qquad
\Omega_{\Lambda}=\frac{2}{3}.$$
There is no "fine tuning" problem in our model
since the ratio of the $\lambda$-part energy density to the mass-energy
density of the
ordinary matter remains during the cosmic evolution constant.

\section{The force exerting on the moving bodies in the cosmic
quaternionic field}
In analogy with the electromagnetic
field, the quaternionic field acts on the moving "charged" objects with
the Lorenz-like force.
In \cite{MK} we supposed that the $\Phi$-field interacts
with all form of energy and matter and the
coupling constant k is from the dimensional reason equal to $\sqrt{G}$.
The "charge" of the $\Phi$-field for a point mass
$m_{0}$ is $\sqrt {G}m_{0}$.
Since the momentum of a moving particle is $p_i=m_0v_i,\quad i=1,2,3,0$,
its
current is given as $J_i= \sqrt{G} m_0v_i=m_{0}\sqrt{G}v_i$. For the
Lorentz-like  force
acting on this particle in the $\Phi$-field we get \cite{X}
$$F_i=c^{-1}\sqrt{G}m_0\Phi v_i=  c^{-1}\sqrt{G}\Phi p_i.\quad\eqno(14)$$

Now, for the sake of simplicity, we confine ourselves to the
non-relativistic case, i.e.
we suppose that $m=$ const. and $v\ll c$. Then Eq.(14) turns out to be
$$ m\dot v =c^{-1}\sqrt{G}\Phi mv. \quad \eqno(15)$$
The quaternionic field affects the following kinetic quantities of the moving
bodies:\\
(i) The velocity of the moving bodies in the presence of
the $\Phi$-field increases. Since $c^{-1}\sqrt{G}\Phi=1/t$ we get the
following simple equation of
equation $\dot v=\beta v $, where $\beta=1/t$,
the solution of which is $v= Ct$.
A free moving object in the quaternionic field
is accelerated by a constant acceleration $C$. This
acceleration is due to the immense smallness of $\beta\approx 1/10^{18}$
in the present-day extremely
small.\\
(ii) The kinetic energy of the moving bodies in $\Phi$-field increases,
too.
The gain of kinetic
energy of a moving body per time unit in the quaternionic field
if $(f_i\parallel v_i)$ is
         $$ {dE\over dt}=F_iv_i = c^{-1}
\sqrt {G} \beta mv^2 =2 \sqrt{G}c^{-1}\Phi E_{kin}=2\beta E_{kin}.$$
Again, the increase of the kinetic energy of a moving object is extremely
small. However, for a rapid rotating dense body it may represent a
considerable value \cite{X}

It is noteworthy that the force exerting on cosmical body in cosmic
quaternionic field is {\it always}
parallel to the direction of the velocity. This means that velocity of
moving bodies in cosmic $\Phi$-field is in all direction
increasing. For example, the moving bodies stemming from a exploding cosmic
body is equally speedup as those falling in the collapsing center.
 
\section{The possible mechanism for the particle creation in the
cosmic quaternionic field}

According to
quantum theory, the vacuum contains many virtual particle-anti-particle
pairs whose lifetime
$\Delta t$ is bounded by the uncertainty relation $\Delta E\Delta t>h$
The proposed mechanism for the particle creation in the $\Phi$-field is
based on the force relation (14).
During the lifetime of the virtual particles
the Lorentz-like force (14) acts on
them and so they gain energy. To estimate this
energy we use simple heuristic arguments.
As is well-known, any virtual particle can only exist within limited
lifetime and its kinetics is bounded to the uncertainty relation
$\Delta p \Delta x>h$.
Therefore, the momentum of a virtual particle $p$ is approximately
given as $p\approx h\Delta x^{-1}$.
If we insert this momentum into Eq.(14) and multiply it by
$\Delta x$, then the energy of virtual
particle $\Delta E$, gained from the ambient $\Phi$-field during
its lifetime, is
$$F\Delta x= \Delta E =\sqrt{G} \Phi(t){h\over c}.\quad \eqno(15)$$
When the $\Phi$-field is sufficiently strong then it can supply enough
energy to the virtual particles during their lifetime and so
spontaneously create real particles
from the virtual pairs. The energy necessary for a particle to be created
is equal to $m_vc^2$ ($m_v$ is the rest mass of the real particle).
At least, this
energy must be supplied from the ambient $\Phi$-field to a virtual particle
during its lifetime. Inserting $\Phi$ into Eq. (15), we have
$$\Delta E\approx {h\over (t+t_0)},$$
Two cases may occur:
(i) If $m_vc^2<\Delta
E$, then the energy
supplied from the $\Phi$-field is sufficient for creating real particles of
mass $m_v$
and, eventually, gives them an additional kinetic energy.
(ii) If $m_v>c^2\Delta
E$, then
the supplied energy is not sufficient for creating the real particles  of
mass $m_v$ but only the energy excitations in vacuum.\\
We see that
there is a plausible mechanism of particle production
in the cosmic $\Phi$-field
This production is very intensive in the early universe
but practically negligible in the present time.

\section{The energy density of the cosmic $\Phi$-field as a possible
candidate for the
black energy}

As has been shown in the previous Chapters,
when taking the field energy density of cosmic quaternionic
field as the vacuum
energy density, the problems presented in the Introduction may be
resolved. The problem of the
cosmological constant, because the value of $\Lambda$ is consistent with data,
the tuning and age problems because
because the ratio of mass to vacuum energy density does not
vary during the cosmical evolution and
the age of the universe is large
enough to evolve the globular cluster.
The flatness problem is also solved because the sum of the black energy
to
the
ordinary matter yields just the critical energy density.
Moreover, in the cosmic quaternionic field,
there exists a plausible mechanism of
the particle creation.
We conclude that the energy density of the $\Phi$-field represent
a relativistic
quantity satisfying Gliner's requirements which is smoothly
distributed in space. It causes the speedup of the universe,
balances the total energy density to the critical one and
gives a plausible mechanism for particle creation.

It is generally believed that dark energy was less
important in the past and will become more important in the future.
In our model the value of black energy is proportional to $1/t^2,$
therefore, its value becomes very large in past and will be adequate
large in the future. We remember that the force exerted by the cosmic
$\Phi$-field on moving bodies acts always in the direction of the
velocity. That means that the high value of the black energy in the
early universe does not interfere with the structure forming, contrarily,
it accelerates it.

The
evolution of the universe with the cosmic quaternionic field
can be briefly sketched as follows:
The cosmic evolution started with purely
field-dominated with the inflation, after which
a massive creation of particles began together with enormous release of
entropy. During the time interval
$(\approx 0,10^{-20})$ the masses of the created particles lei in the
range from $10^{-5}$ to $10^{-27}$ g.
Their kinetic energy
was $E_{kin}=[h/(t+t_0)]-m_0c^2$.
$E_{kin}$ of the created nucleons
has reached values up to $10^{-5}$ erg, which corresponds to
the temperature of $10^{21}$ K.
Today, energies of the virtual pairs, gained during their lifetime,
are immense small, therefore, it represents only a certain local energy
excitations of the vacuum.
The large vacuum energy density of the cosmic quaternionic field at the
early stage of the universe
accelerates its structure forming. From what has been said above
we conclude that the energy density of the cosmic
quaternionic field
might be a possible candidate for the black energy because (i) it has the
 value consistent with data (ii) it does not suffer from the cosmological
 constant, fine-tuning, age and
flatness problems (iii) it yields a plausible mechanism particle production
and (iv) it accelerated the structure forming in the early universe.

\end{document}